\documentclass[aps,amsmath,amssymb,twocolumn,prl,superscriptaddress, 10pt]{revtex4-2}

\usepackage{graphicx,tikz,amsfonts}% Include figure files
\usetikzlibrary{arrows.meta, decorations.pathreplacing}
\usepackage{dcolumn}% Align table columns on decimal point
\usepackage{bm}% bold math
\usepackage{epsfig,amsthm}
\usepackage{color,fancybox}
\usepackage{colordvi}
\usepackage{relsize}
\usepackage{accents}
\usepackage{wasysym} 
\usepackage{enumitem}
\usepackage{mathtools,slashed}
\usepackage{times}
\usepackage{newtxmath}
\usepackage{anyfontsize}
\RequirePackage{fix-cm}
\usepackage{subcaption}

\usepackage{hyperref}
\usepackage[capitalize]{cleveref}

\hypersetup{
	colorlinks=true,
	linkcolor=CitingColor,
	citecolor=CitingColor,
	urlcolor=CitingColor
}

\hypersetup{
	colorlinks=true,
	linkcolor=CitingColor,
	citecolor=CitingColor,
	urlcolor=CitingColor
}

\DeclareMathAlphabet\mathbfcal{OMS}{cmsy}{b}{n}

\newcommand{\ket}[1]{\ensuremath{|#1\rangle}}
\newcommand{\bra}[1]{\ensuremath{\langle #1|}}

\newcommand{\proj}[1]{\ket{#1}\!\bra{#1}}
\newcommand{\be}{\begin{equation}}
\newcommand{\ee}{\end{equation}}
\newcommand{\ba}{\begin{eqnarray}}
\newcommand{\ea}{\end{eqnarray}}

\newcommand{\tr}{{\rm tr}}

\newcommand{\id}{\mathbb{I}}

\newtheorem{result}{Result}
\newtheorem{question}{Question}

\newtheorem{observation}{Observation}

\definecolor{nred}{rgb}{0.9,0.1,0.1}
\definecolor{nblack}{rgb}{0,0,0}
\definecolor{nblue}{rgb}{0.2,0.2,0.8}
\definecolor{ngreen}{rgb}{0.2,0.5,0.2}
\definecolor{ublue}{rgb}{0,0,0.5}

\definecolor{pur}{rgb}{0.75,0,0.75}
\definecolor{nngrn}{rgb}{0,0.5,0.5}
\definecolor{CitingColor}{rgb}{0,0.3,1}

\begin{document}
\title{Bell nonlocality from compatibility of entanglement-breaking channels}

\author{Gelo Noel M. Tabia}
\email{gelo.tabia@gmail.com}
\affiliation{Hon Hai (Foxconn) Quantum Computing Research Center, Taipei 114, Taiwan}
\affiliation{Department of Physics and Center for Quantum Frontiers of Research \& Technology (QFort), National Cheng Kung University, Tainan 701, Taiwan}

\author{Chung-Yun Hsieh}
\email{chung-yun.hsieh@bristol.ac.uk}
\affiliation{H. H. Wills Physics Laboratory, University of Bristol, Tyndall Avenue, Bristol, BS8 1TL, United Kingdom}
\affiliation{ICFO - Institut de Ci\`encies Fot\`oniques, The Barcelona Institute of Science and Technology, 08860 Castelldefels, Spain}

\author{Min-Hsiu Hsieh}
\email{min-hsiu.hsieh@foxconn.com}
\affiliation{Hon Hai (Foxconn) Quantum Computing Research Center, Taipei 114, Taiwan}
 
\date{\today}

\begin{abstract}

Locally classical behavior is often interpreted as evidence that a global classical description should exist. In particular, when individual processes admit classical (measure-and-prepare) realizations, it is natural to expect at least one compatible joint implementation that remains classical.
We show that this intuition fails in a simple broadcast setting with one input system and two outputs. Specifically, we construct pairs of compatible channels that are entanglement-breaking individually and thus admit classical descriptions,
yet {\em every} joint broadcast realization is necessarily Bell-nonlocal between the outputs, even for a maximally mixed input.
This establishes a form of compatibility-induced activation of nonlocality: Bell nonlocality arises not from entanglement in the inputs or marginal channels, but solely from the requirement that these marginals admit a common global implementation. In this sense, compatibility is not merely a consistency condition on local descriptions, but a mechanism that can enforce nonclassical correlations at the global level.

\end{abstract}

\maketitle

% \noindent{\bf\em Introduction---}
% \noindent{\bf INTRODUCTION}
\section{INTRODUCTION}
% \noindent 
Entanglement is undoubtedly a central resource underpinning various aspects of quantum science and technologies~\cite{HorodeckiRMP}, such as quantum teleportation~\cite{Bennett93}, super-dense coding~\cite{Bennett1992}, quantum communication~\cite{wilde_2017,DattaITIT2013,AnshuITIT2019}, and device-independent protocols based on Bell nonlocality~\cite{Brunner2014RMP,Bell1964,Wiseman2007PRL,Jones2007PRA,UolaRMP2020,Xiang2022PRXQuantum,Cavalcanti2016PRA,Nery2020,Quintino2016,Hsieh2016PRA,Cavalcanti2016,Ku2022NC,Ku2024,Hsieh2024,Hsieh2024-2,Hsieh2024-3,Branciard2012}. In particular, Bell tests provide a standard method for certifying nonclassical correlations associated with entanglement~\cite{Brunner2014RMP,Bell1964}.

A recurring theme in quantum theory is whether nonclassical behavior can emerge from structures that appear entirely classical when viewed locally. More broadly, one may ask whether locally classical behavior can always be combined into a globally classical description.

To investigate this question, we consider a broadcasting scenario with one input system and two outputs. Two channels ${A\to B}$ and ${A\to C}$ are said to be compatible~\cite{Haapasalo2021Quantum,Hsieh2022PRR} if they arise as marginals of a common broadcast channel ${A\to BC}$. Compatibility constraints determine what type of global correlations can arise from locally specified processes, which motivates studying how quantum resources such as entanglement and Bell nonlocality manifest in compatible channel structures.

In classical theory, compatible stochastic maps always admit a joint realization within a single probabilistic model (see, e.g., Eq.~(18) in Ref.~\cite{Hsieh2022PRR}). In quantum theory, entanglement-breaking channels~\cite{Horodecki2003} provide the closest analogue, as they allow a classical simulation via measure-and-prepare implementations~\cite{Horodecki2003}, and generate only classical correlations with external systems, which implies zero quantum capacity~\cite{Shor2002EB}. 
This suggests that consistency conditions among such classically realizable channels are not expected to enforce genuinely nonclassical global behavior, and that at least one broadcast realization should admit a classical description.

Given two entanglement-breaking marginal channels \mbox{$A\to B$} and $A\to C$, we therefore ask whether there always exists at least one joint broadcast extension $A\to BC$ that remains separable across $BC$. One might expect the answer to be affirmative, especially in the absence of additional quantum resources (see Fig.~\ref{Fig:main_question} for an illustrative example).

Remarkably, there exist pairs of compatible entanglement-breaking channels whose joint broadcast extensions are {\em all} Bell-nonlocal.
In particular, every such extension violates the CHSH inequality for the state between the remote outputs $B$ and $C$, even when the input in $A$ is maximally mixed. 
This rules out any explanation based on coherence or entanglement supplied by the input, demonstrating that Bell nonlocality can arise purely from the compatibility constraints imposed on otherwise classically realizable channels.

\begin{figure}[t]
\scalebox{0.73}{

\begin{tikzpicture}[
    arrow/.style={-{Stealth[length=12pt, width=8pt]}, line width=4pt, draw=blue!70!black},
    label/.style={text=blue!70!black, font=\large\bfseries}
]

% Left vertex label
\node[label] at (-0.5, 0) {$\dfrac{\mathbb{I}}{2}$};

% Origin point of arrows
\coordinate (O) at (0.4, 0);

% Two arrow tips
\coordinate (A) at (5.5,  1.8);
\coordinate (B) at (5.5, -1.8);

% Top arrow with label (placed well above the arrow)
\draw[arrow] (O) -- (A);
\node[label, above] at (2, 1.6) {Entanglement-breaking};

% Bottom arrow with label (placed well below the arrow)
\draw[arrow] (O) -- (B);
\node[label, below] at (2, -1.6) {Entanglement-breaking};

% Cyan brace on the right
\draw[line width=3pt, draw=cyan!80!blue, decorate,
    decoration={brace, amplitude=10pt}]
    (5.7, 2.1) -- (5.7, -2.1);

% Bracket label
\node[text=cyan!80!blue, font=\large\bfseries, align=left] at (8.2, 0)
    {Can this always\\ be separable?};

\end{tikzpicture}

}
\caption{
For two entanglement-breaking marginal channels $A \to B$ and $A \to C$, we ask whether they can always be broadcast such that the joint output on $BC$ is separable. For example, take state-preparation channels that output a  maximally mixed qubit. These are entanglement-breaking, and can be jointly realized by the state-preparation channel $(\cdot)_A \mapsto \mathbb{I}_{BC}/4$, whose output is always separable.
}
\label{Fig:main_question}
\end{figure}

More precisely, we explicitly construct two entanglement-breaking channels with the same input system ($A$) and different output systems ($B$ and $C$, respectively) such that: 
\begin{enumerate}[label=(\roman*)]
    \item 
    They are {\em (broadcast) compatible}~\cite{Haapasalo2021Quantum,Hsieh2022PRR}, i.e., there exists a broadcast channel from $A$ to $BC$ (their so-called {\em joint broadcast extension}) that can simultaneously realize them via partial trace. 
    \item
    {\em All possible} broadcast channels realizing them generate an entangled state in $BC$ that violates the {\em Clauser-Horne-Shimony-Holt} (CHSH) inequality~\cite{CHSH} for a maximally mixed input in $A$.
\end{enumerate}

Together, these observations show that the resulting nonlocality in $BC$ cannot be attributed to locally accessible quantum resources of the marginal channels or of the input system. Rather, it is enforced entirely by compatibility constraints on the marginals.
Viewed operationally, this means that compatibility acts not merely as a restriction on which broadcast realizations are possible, but also as a mechanism that leaves only extensions that exhibit nonlocality. 

Our results bear conceptual similarities to activation phenomena~\cite{Popescu1995PRL, Navascues2011PRL, Palazuelos2012PRL, Villegas-Aguilar2024,Quintino2016,Hsieh2016PRA,Cavalcanti2013,Masanes2008PRL,Liang2012PRA, Zhang_2020_channel_chsh,Hsieh2020} for quantum channels, in particular, the activation of CHSH nonlocality from CHSH-breaking channels. 
More precisely, Ref.~\cite{Zhang_2020_channel_chsh} showed that there exists an amplitude-damping channel whose outputs can never violate the CHSH inequality, yet two appropriately combined copies of the channel can produce CHSH-nonlocal states. 
Likewise, Ref.~\cite{Hsieh2020} presents a bipartite channel whose output cannot violate any Bell inequality, yet sufficiently many copies can generate nonlocality.
In those scenarios, the nonlocality is activated only after additional operational structure is introduced, such as tensor-product composition or tailored inputs and measurements. By contrast, here nonlocality is already enforced at the level of the marginal constraints and does not require any additional copies, processing steps, or auxiliary resources beyond compatibility.
% \\

% \noindent{\bf RESULTS}
\section{RESULTS}
% \noindent 
For our argument, we recast the compatibility problem as a constrained extension problem for Choi states. First, we construct compatible entanglement-breaking marginals from a tripartite state with a maximally mixed input marginal. Second, we show that the broadcast channel it represents (through the Choi state~\cite{Choi1975,Jamiolkowski1972}) yields a CHSH violation. Third, using a semi-definite program (SDP)~\cite{BoydVandenberghe2004,SkrzypczykCavalcanti2023},
we prove that every global extension must remain CHSH nonlocal.
% \\

\subsection{Preliminaries}
% \noindent{\bf\em Preliminaries---}
First, we formalize the notion of compatibility used in this work. Consider three finite-dimensional systems $A$, $B$, and $C$.
A channel (i.e., completely positive trace-preserving linear map) from  $A$ to the bipartite system $BC$, denoted by $\mathcal{G}_{A\to BC}$, is called a {\em broadcast channel} (here, the subscripts denote the system(s) on which the object lives or acts on).
For a pair of channels $\mathcal{N}_{A\to B}$ and $\mathcal{N}_{A\to C}$, 
a fundamental question is: {\em Can they be simultaneously realized by a broadcast channel from $A$ to $BC$?}
This asks for a broadcast channel $\mathcal{G}_{A\to BC}$ achieving
\begin{align}\label{Eq:broadcast def}
{\rm tr}_C\circ\mathcal{G}_{A\to BC} = \mathcal{N}_{A\to B}\quad\&\quad{\rm tr}_B\circ\mathcal{G}_{A\to BC} = \mathcal{N}_{A\to C}.
\end{align}
If such a broadcast channel exists, $\mathcal{N}_{A\to B}$ and $\mathcal{N}_{A\to C}$ are said to be {\em (broadcast) compatible}~\cite{Haapasalo2021Quantum,Hsieh2022PRR}.
We call $\mathcal{G}_{A\to BC}$ in Eq.~\eqref{Eq:broadcast def} a {\em broadcast realization} of $\mathcal{N}_{A\to B}$ and $\mathcal{N}_{A\to C}$.

In general, a broadcast realization is not unique.
For instance, when $B,C$ are qubits, the channels $\mathcal{N}_{A\to B}(\cdot) = {\rm tr}(\cdot)\id_B/2$ and $\mathcal{N}_{A\to C}(\cdot) = {\rm tr}(\cdot)\id_C/2$ are compatible, and both $(\cdot)_A\mapsto\id_{BC}/4$ and $(\cdot)_A\mapsto\proj{\Phi^+}_{BC}$,where $
\ket{\Phi^+}_{BC}\coloneqq(\ket{00}+\ket{11})_{BC}/\sqrt{2}$, are admissible broadcast realizations (see also Fig.~\ref{Fig:main_question}).
Importantly, not all pairs of channels are compatible.
For instance, the identity channel from $A$ to $B$ is not compatible with the identity channel from $A$ to $C$---otherwise, the no-cloning theorem~\cite{Wootters1982,Dieks1982} would be violated.
This is the so-called {\em (broadcast) incompatibility}, a quantum phenomenon generalizing the physics of the no-cloning theorem.

A quantum channel $\mathcal{N}_{A\to B}$ is {\em entanglement-breaking}~\cite{Horodecki2003} if 
$
(\mathcal{N}_{A\to B}\otimes\mathcal{I}_{C})(\eta_{AC})
$
is separable for every bipartite state $\eta_{AC}$ and finite-dimensional system $C$, where $\mathcal{I}_{C}$ is the identity channel acting on $C$.
Such channels break entanglement between their outputs and any external system.
A useful characterization of entanglement-breaking channels is via the so-called Choi state~\cite{Choi1975,Jamiolkowski1972}.
Formally, the channel $\mathcal{N}_{A\to B}$'s {\em Choi state} is the bipartite state in $AB$ defined by (note that $A,B$ can have different dimensions)
\begin{align}\label{Eq:Choi state def}
\mathcal{J}_{AB}^{(\mathcal{N})}\coloneqq({\mathcal{I}_{A}}\otimes\mathcal{N}_{A\to B})(\proj{\Phi^+}_{AA}),
\end{align}
where $\ket{\Phi^+}_{AA}\coloneqq\sum_{n=0}^{d-1}\ket{nn}_{AA}/\sqrt{d}$ is maximally entangled in the bipartite system consisting of two identical copies of $A$ ($d$ denotes the dimension of $A$).
Moreover, $\mathcal{N}_{A\to B}$ is entanglement-breaking {\em if and only if} its Choi state $\mathcal{J}_{AB}^{(\mathcal{N})}$ is separable~\cite{Horodecki2003,Holevo2008}.
Through the Choi-Jamiolkowski isomorphism, compatibility of channels can be reformulated as marginal constraints on the corresponding tripartite Choi states.

Now, a bipartite quantum state $\rho_{AB}$ is said to be {\em Bell-nonlocal}, or {\em nonlocal} for short, if there exists a choice of single-party measurements $\{A_{a}^{x}\}_{a,x}$ and $\{B_{b}^{y}\}_{b,y}$, i.e., for each $x,y$, both $\{A_{a}^{x}\}_{a}$ and $\{B_{b}^{y}\}_{b}$ are positive operator-valued measures (POVMs)~\cite{QIC-book}, such that the quantum correlation $\vec{P}\coloneqq\{P(a,b|x,y)\}_{a,b,x,y}$ given by
\begin{equation}
    P(a,b|x,y) = \mathrm{tr}\left(\rho_{AB} {(A_{a}^{x} \otimes B_{b}^{y})}\right)\quad{\forall\,a,b,x,y}
\end{equation}
can violate some Bell inequality (namely, this correlation is nonlocal). 
For any two-qubit state $\rho_{AB}$, Horodecki {\em et al.}~\cite{horodecki2000PLA} showed that maximum CHSH value achievable with $\rho_{AB}$, denoted by $\mathrm{CHSH}(\rho_{AB})$, can be computed from its Pauli correlation matrix $T$ given by
$T_{ij} = \mathrm{tr}(\rho_{AB}(\sigma_{i,A} \otimes \sigma_{j,B})),$
% \end{equation}
where $\sigma_i$'s are Pauli $X$, $Y$, $Z$ matrices for $i=1,2,3$, respectively. 
Consider the singular value decomposition of $T$ as $T = U\Sigma V^T$ with $\Sigma = \mathrm{diag}(t_{1},t_{2}
,t_{3})$, and let
$\vec{u}_i$ and $\vec{v}_i$ denote the columns of $U$ and $V$, respectively.
Let $t_{1}$ and $t_{2}$ be the two largest singular values of $T$.
Then, we have that~\cite{horodecki2000PLA}
\begin{equation}
\label{eq:chshValueFromT}
    \mathrm{CHSH}(\rho_{AB}) = 2\sqrt{t_{1}^2 + t_{2}^2}.
\end{equation}
The CHSH observables that attain this optimal value are given by $A_{i} = \vec{a}_{i}\cdot\vec{\sigma}$ and $B_{j} = \vec{b}_{j}\cdot\vec{\sigma}$ for $i,j=1,2$, where $\vec{a}_i=\vec{u}_i$, $\vec{b}_{j} = (\vec{v}_1 -(-1)^{j} \vec{v}_2)/\sqrt{2}$, and $\vec{\sigma}\coloneqq(\sigma_1,\sigma_2,\sigma_3)$.

With these observables, we can evaluate the CHSH value with the corresponding CHSH Bell operator defined by
\begin{equation}
\label{eq:chshBelloperator}
    \beta^{\mathrm{CHSH}} := A_{1}\otimes B_{1} + A_{1}\otimes B_{2} + A_{2}\otimes B_{1} - A_{2}\otimes B_{2}.
\end{equation}

\subsection{Activating nonlocality from compatibility of entanglement-breaking channels}
% \noindent{\bf\em Activating nonlocality from compatibility of entanglement-breaking channels---}
In a broadcast scenario, there exist entanglement-breaking channels $\mathcal{E}_{A\to B}$ and $\mathcal{E}_{A\to C}$ such that every joint broadcast extension from $A$ to $BC$ inevitably produces CHSH-nonlocal correlations in $BC$. Thus, while the marginal channels individually admit only classical behavior, any compatible broadcast realization is necessarily nonlocal.

To prove the claim, we provide an explicit example in a three-qubit system ($ABC$); details are provided in Appendix A.
% Methods.
It is a rank-2 state:
% Example:
\begin{equation}\label{Eq:3-qubit example}
    \rho_{ABC} \coloneqq  \sum_{i=1}^{2} \lambda_{i} \proj{\psi_{i}}_{ABC} 
\end{equation}
and we take its bipartite marginals in $AB$ and $AC$:
\begin{align}\label{Eq:examples}
\rho_{AB}\coloneqq{\rm tr}_C(\rho_{ABC})\quad\&\quad
\rho_{AC}\coloneqq{\rm tr}_B(\rho_{ABC}).
\end{align}
It can be verified that $\rho_{A} = \id/2$ and that both bipartite marginals have positive partial transpose (PPT)~\cite{Peres1996PRL,Horodecki1996PLA} and are thus separable. By viewing them as Choi states and using the fact that a Choi state is separable if and only if the corresponding channel is entanglement-breaking~\cite{Horodecki2003,Holevo2008}, we conclude that
% \begin{observation}
both $\rho_{AB}$ and $\rho_{AC}$ can be seen as entanglement-breaking channels, and $\rho_{ABC}$ is one of their global extensions.
% \end{observation}

Next, we take $\rho_{BC}$ from the joint extension and show that it violates the CHSH inequality by using Eqs.~(\ref{eq:chshValueFromT}-\ref{eq:chshBelloperator}), which provide us with not just the optimal CHSH value, but also the corresponding CHSH observables and Bell operator $\beta_{BC}^{\mathrm{CHSH}}$.

Finally, to determine whether CHSH nonlocality is unavoidable, we minimize the expectation value of the CHSH Bell operator over all global states compatible with the fixed marginals. Using the method first reported in Ref.~\cite{Hsieh2024resourcemarginal}, we use $\beta_{BC}^{\mathrm{CHSH}}$ as a linear witness for nonlocality by performing
% More precisely, we perform 
\begin{align}
\label{Eq: CHSHtransitivity}
\nonumber \gamma_\star :=
&\min_{\eta_{ABC}}\qquad\qquad \tr\left(\eta_{BC}\beta_{BC}^{\mathrm{CHSH}}\right) \\
&\text{subj. to }  \quad
 \tr_{C}(\eta_{ABC}) = \rho_{AB},\,\,  
  % \nonumber\\  &\qquad \qquad \,\,\, 
 \tr_{B}(\eta_{ABC}) = \rho_{AC},\,\, \nonumber\\ 
 &\qquad\qquad\,\, \eta_{ABC} \succeq 0.
\end{align}
We find that $\gamma_\star > 2$, implying CHSH violation {\em for any compatible global state}. Hence: 

\begin{result}\label{Result 1}
For every three-qubit state $\eta_{ABC}$ satisfying $\rho_{AB}={\rm tr}_C(\eta_{ABC})$ and $\rho_{AC}={\rm tr}_B(\eta_{ABC})$, the state ${\rm tr}_A(\eta_{ABC})$ must exhibit CHSH nonlocality in $BC$. 
\end{result}
 
Because we are using a fixed Bell operator, this means that for every compatible global state $\eta_{ABC}$, we obtain a marginal state $\eta_{BC}$ that violates the CHSH inequality even for the same set of CHSH observables.
Thus, $\gamma_\star$ provides a tight lower bound on the CHSH violation.
Finally, recall that both $\rho_{AB}$ and $\rho_{AC}$ represent some entanglement-breaking channels (say, $\mathcal{E}_{A\to B}$ and $\mathcal{E}_{A\to C}$), and 
every broadcast channel $A\to BC$ realizing them (say, $\mathcal{G}_{A\to BC}$) must have a Choi state $\eta_{ABC}$ that satisfies $\rho_{AB}={\rm tr}_C(\eta_{ABC})$, $\rho_{AC}={\rm tr}_B(\eta_{ABC})$, and ${\rm tr}_A(\eta_{ABC})=\mathcal{G}_{A\to BC}(\id/2)$.
% every $\eta_{ABC}$ in Result~\ref{Result 1} can be viewed as the Choi state of a broadcast channel $A\to BC$.
Thus, Result~\ref{Result 1} implies that {\em all} broadcast channels realizing $\mathcal{E}_{A\to B}$ and $\mathcal{E}_{A\to C}$ must output CHSH-nonlocal states in $BC$.
% \\

\subsection{Physical implications of Result~\ref{Result 1}}
% \noindent{\bf\em Physical implications of Result~\ref{Result 1}---}
Apart from showing that broadcasting two entanglement-breaking channels can guarantee nonlocality, Result~\ref{Result 1} has several other implications, as we now discuss.

First, Result~\ref{Result 1} has a stronger consequence than the unavoidable CHSH nonlocality of the output state. For every joint broadcast extension represented by $\eta_{ABC}$, the bipartition $A|BC$ is entangled. This means that every broadcast channel $A\to BC$ realizing the entanglement-breaking marginals is itself non-entanglement-breaking (see 
Appendix B).
% Methods). 
Equivalently, when acting on half of an entangled state shared with an external reference system $R$, any such broadcast channel can preserve entanglement across the bipartition $R|BC$. Viewing non-entanglement-breaking channels as quantum memory resources~\cite{Rosset2018PRX}, this may be interpreted as an activation of quantum memory induced purely by consistency constraints.

Another implication of Result~\ref{Result 1} is that it provides an instance of nonlocality meta-transitivity, a notion that is strictly stronger than nonlocality transitivity introduced in Ref.~\cite{Chen2026}. More precisely, we say that two bipartite separable states $\rho_{AB}$ and $\rho_{AC}$ exhibit {\em nonlocality meta-transitivity} if every tripartite state $\eta_{ABC}$ compatible with them via partial trace must have a nonlocal $BC$ marginal, i.e., $\tr_A(\eta_{ABC})$ violates a Bell inequality. Result~\ref{Result 1} provides such an example.
Thus, even separable marginals can generate unavoidable nonlocality in the remaining subsystem, provided they are jointly compatible in a suitable way.
% \\

\subsection{A family exhibiting entanglement meta-transitivity}
% \noindent{\bf\em A family exhibiting entanglement meta-transitivity---}
An immediate corollary for our example is that there exists a pair of entanglement-breaking marginal channels such that every compatible global broadcast channel outputs an entangled state when given a maximally mixed input qubit, since entanglement is a prerequisite for nonlocality.

While our CHSH meta-transitivity example is currently obtained numerically, the following one-parameter family demonstrates that the underlying transitivity mechanism is not isolated and admits an analytic characterization at the level of entanglement. This suggests that the compatibility mechanism underlying Result 1 is not a fine-tuned numerical artifact.
Consider
\begin{align}
 \label{Eq: three qubit psi2}
\ket{\psi}_{ABC} &= \sqrt{\tfrac{3}{20}}(\ket{000}+\ket{111}) + \sqrt{\tfrac{5}{20}}(\ket{001} +\ket{110}) \nonumber \\
    &\quad + \sqrt{\tfrac{2}{20}}(\ket{010}+\ket{101}),\\
\ket{\varphi(x)}_{ABC} &= x (\ket{000} +\ket{001}-\ket{110}-\ket{111}) \nonumber \\
&\quad + \sqrt{\tfrac{1}{4}-x^2}
(\ket{010}+\ket{011}-\ket{100}-\ket{101}), \end{align}
where $0\le x \le \tfrac{1}{2}$, and take the mixture 
\begin{align}
\rho_{ABC}(x)\coloneqq \Big(p(x)\proj{\psi} + [1-p(x)]\proj{\varphi(x)}\Big)_{ABC}
\end{align}
with
$
p(x)\coloneqq\left(20 x \sqrt{1-4 x^2}\right)/\left(20 x \sqrt{1-4 x^2}+\sqrt{10}\right).
$
Then, for $\rho_{ABC}(x)$ to become a valid Choi state of a broadcast channel, one can choose
\begin{equation}
\label{Eq:parameterX}
x = \tfrac{4}{9} + q\left(\sqrt{(\sqrt{15}+5)/40}-\tfrac{4}{9}\right), \quad 0\le q\le 1.
\end{equation}
The relevant one-parameter family of channels is given by the $AB$ and $AC$ marginals of $\rho_{ABC}(x)$. It is straightforward to check that when we choose $x$ according to Eq.~(\ref{Eq:parameterX}), then $\rho_{A}$ is maximally mixed, and that $\rho_{AB}(x)$ and $\rho_{AC}(x)$ are PPT and thus separable. 

 For every such pair of separable states, we can show that they exhibit {\em entanglement meta-transitivity} (namely, they are separable and every tripartite state in $ABC$ compatible with them must have an entangled $BC$ marginal) by solving the optimization problem~\cite{Tabia2022}:
\begin{align}
\label{Eq: NPTtransitivitySDP}
\nonumber \lambda_\star :=
&\max_{\eta_{ABC}}\qquad\qquad \lambda  \\
&\text{subj. to }  \quad
 \tr_{C}(\eta_{ABC}) = \rho_{AB},\,\,  
  % \nonumber\\  &\qquad \qquad \,\,\, 
 \tr_{B}(\eta_{ABC}) = \rho_{AC},\,\, \nonumber\\ 
 &\qquad\qquad\,\, \eta_{ABC} \succeq 0, \quad\eta_{BC}^{\Gamma_B} \succeq  \lambda \mathbb{I}, 
\end{align}
where $\eta_{BC} = \tr_{A}(\eta_{ABC})$ and $\Gamma_B$ denotes the partial transpose in $B$.
The optimal solution is the global state $\eta_{ABC}$ compatible with $\rho_{AB}$ and $\rho_{AC}$ such that the partial transpose of its marginal state in $BC$ has the maximal smallest eigenvalue $\lambda_\star$.
If this ``worst-case'' $\lambda_\star < 0$, it implies that every compatible global state has a $BC$ marginal state with negative partial transpose
% state in $BC$ 
and therefore $\eta_{BC}$ is always entangled.\\

\begin{figure*}[t!]
    \centering
    \begin{subfigure}[t]{0.5\textwidth}
        \centering
        \includegraphics[width=0.7\linewidth]{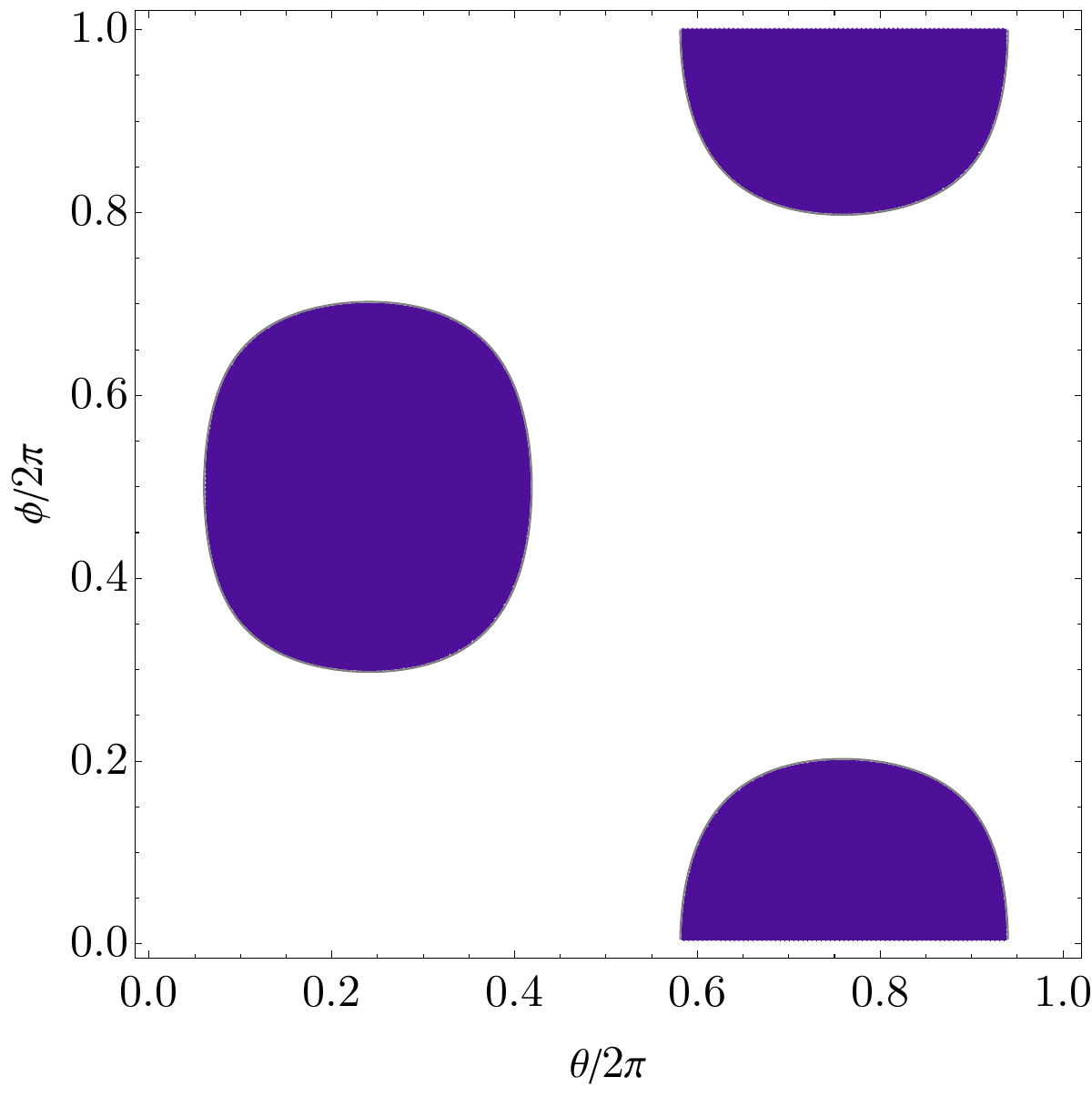}
        \caption{}
    \end{subfigure}%
    ~ 
    \begin{subfigure}[t]{0.5\textwidth}
        \centering
        \includegraphics[width=0.8\linewidth]{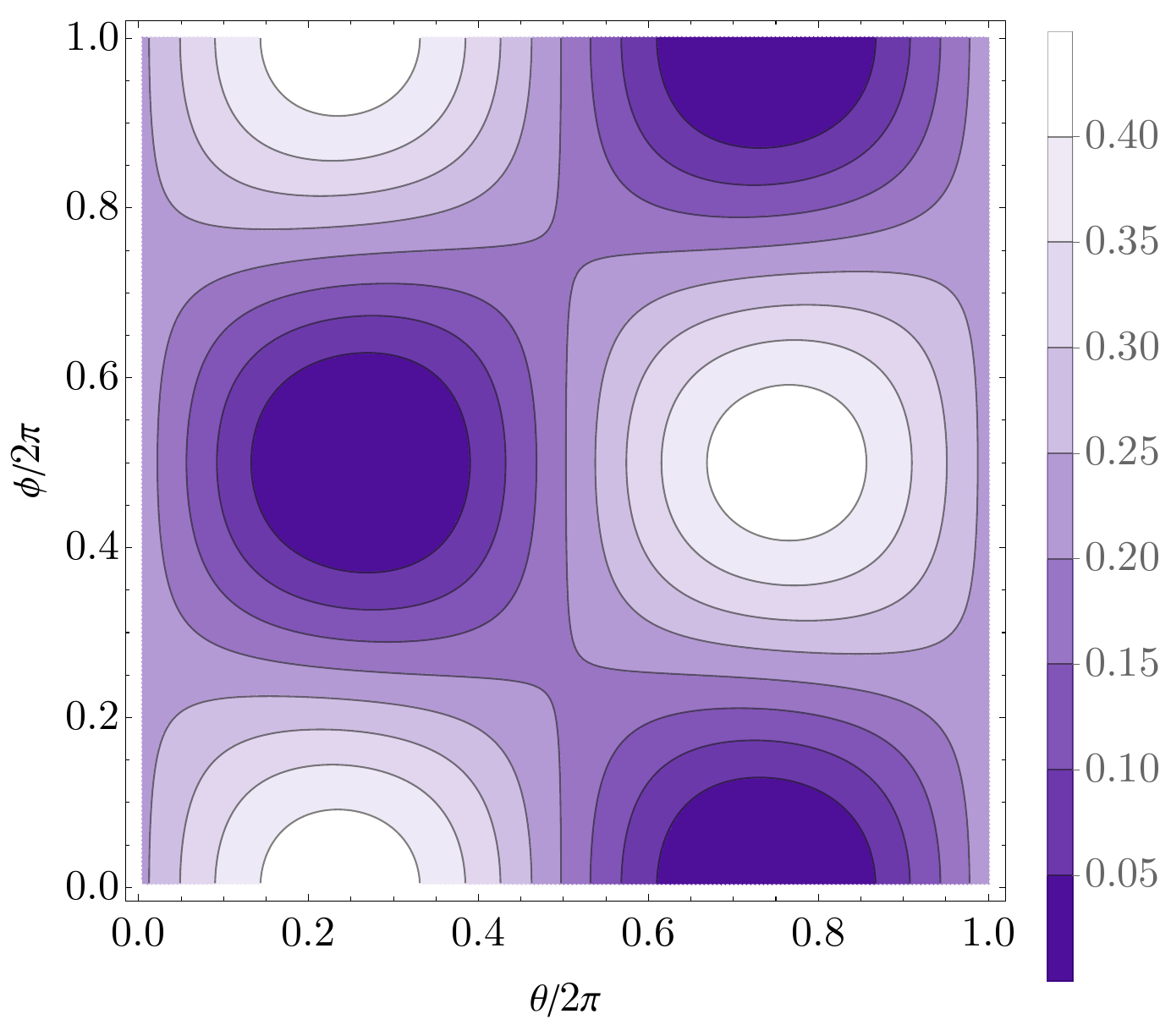}
        \caption{}
    \end{subfigure}
    \caption{Contour plots for the (a) CHSH value and (b) negativity for the output of the broadcast channel corresponding to Eq.~(\ref{Eq:3-qubit example}) for pure input state with Bloch vector $\vec{r} = (\sin\theta\cos\phi, \sin\theta\sin\phi,\cos\theta)$. For (a), CHSH is violated in the white region, which represents roughly two-thirds of the Haar-sampled points. For (b), the negativity is more than $10^{-5}$ for all  sampled points.}
\label{Fig:contourPlotsGenericInput}
\end{figure*}

\subsection{Quantum correlations beyond the maximally mixed input}
% \noindent{\bf\em Quantum correlations beyond the maximally mixed input---}
Our result shows that for the entanglement-breaking channels $\mathcal{E}_{A\to B}$ and $\mathcal{E}_{A\to C}$ associated with $\rho_{AB}$ and $\rho_{AC}$, any joint broadcast extension $\mathcal{G}_{A\to BC}$ produces a CHSH-nonlocal state in $BC$ for a maximally mixed input. 

One may then ask what happens for other inputs to the broadcast channel. 
This question is particularly relevant because the maximally mixed input could, a priori, represent a highly special operating point of the construction.

Numerically, we find that
the output remains entangled for Haar-random pure-state inputs, while CHSH violation occurs only for certain inputs (cf. Fig.~\ref{Fig:contourPlotsGenericInput}). 
Thus, although Bell nonlocality is confined to specific regions of the input space, entanglement appears to remain robust in the output of the channel despite its entanglement-breaking marginals.
To be more specific (see also 
Appendix C):
% Methods):
\begin{observation}
For the example in Eq.~(\ref{Eq:3-qubit example}), numerical sampling of $10^6$ Haar-random pure inputs revealed no separable outputs within numerical precision ($10^{-8}$), and CHSH violation occurs for approximately two-thirds of the sampled inputs.
\end{observation}
The maximally mixed input therefore does not appear to be a singular point of the construction, and output entanglement may be substantially more generic than the CHSH violation itself.
Thus, while CHSH nonlocality occurs only for a subset of inputs for the example in Eq.~(\ref{Eq:3-qubit example}), output entanglement appears to be a generic feature of this construction.
% \\

% \noindent{\bf\em Discussion---}
% \noindent{\bf DISCUSSION}
\section{DISCUSSION}
% \noindent 
Here we have shown that compatibility constraints alone can enforce Bell nonlocality. Although each marginal channel in our construction is entanglement-breaking and admits a classical measure-and-prepare realization, every joint broadcast extension necessarily generates CHSH-nonlocal states.

This provides a form of compatibility-induced activation of nonlocality. Unlike conventional activation phenomena, where nonlocality is revealed only after tensor-product composition, collective processing, specially chosen inputs, or the introduction of additional operational resources, we illustrate that it can follow solely from the requirement that the marginal channels admit a common broadcast realization. No extra copies, ancillary systems, filtering procedures, or collective measurements are required. The activation is therefore enforced entirely by consistency constraints on the global realization.

Equivalently, by viewing the marginal channels through their Choi states, the result admits a purely state-theoretic reformulation in terms of marginal constraints.
This perspective highlights that the phenomenon can be understood entirely at the level of bipartite states subject to consistency conditions. It can thus be interpreted as a manifestation of nonlocality meta-transitivity. As an application, our findings show that the compatibility of two separable states can ensure nonlocality.

Note that if we merely seek a three-qubit state where (i) $AB$ and $AC$ are separable, (ii) $A$ is maximally mixed, and (iii) $BC$ violates CHSH, then a simple construction is
$ \rho_{ABC}(p) = (1/2)\left(\proj{0}\otimes\rho_{BC}(p) + 
    \proj{1}\otimes{\id/4}\right)$
for $p>1/\sqrt{2}$, where
$\rho_{BC}(p) = p\proj{{\Phi^+}} + (1-p){\id/4}$.
However, these marginals place essentially no restrictions on the compatible global state and therefore admit both nonlocal and fully classical compatible extensions (see Fig.~\ref{Fig:main_question}). 
The challenge is therefore not to produce a state with separable $AB$, separable $AC$, and nonlocal $BC$, but to make the nonlocality unavoidable across all compatible extensions.

By contrast, the present construction is stronger than the one above: the same marginal constraints force {\em every} compatible global realization to exhibit CHSH nonlocality in $BC$.
This is technically more demanding than certifying entanglement transitivity, since CHSH violation depends nonlinearly on the two-qubit correlation matrix and thus makes it harder to guarantee across the full set of compatible extensions.

Together with the one-parameter family and the broad region of entangled outputs observed under generic inputs, these findings suggest that this compatibility-induced nonclassicality is not confined to a special isolated construction.

These observations suggest a few directions for future work. More broadly, our results indicate that compatibility can be viewed as a structural resource rather than merely a consistency condition on global realizations. It may therefore provide a framework for understanding transitivity phenomena of quantum resources in networked and multipartite scenarios, and for identifying situations in which global nonclassicality is enforced by otherwise classical local descriptions.
% \\

% \noindent{\bf METHODS}
\section{Appendix}
\subsection{Appendix A: Details for Result~\ref{Result 1}}
% \noindent{\bf\em Details for Result~\ref{Result 1}---}
To prove {Result~\ref{Result 1}}, we provide an explicit example in a three-qubit system ($ABC$).
We start with the state in Eq.~\eqref{Eq:3-qubit example}, 
$ \rho_{ABC} \coloneqq  \sum_{i=1}^{2} \lambda_{i} \proj{\psi_{i}}_{ABC} $,
where
\begin{align}
    \lambda_1 &= 0.473317526432882, \nonumber \\
    \lambda_2 &=     0.526682483567118, 
\end{align}
and
\begin{align}
    \ket{\psi_1}_{ABC} &=
    \begin{pmatrix}
   -0.209999952108617 \\
   0.174094868773554 \\
   0.117224725538884 \\
   0.182754562207009 \\
   0.338480049085443 \\
  -0.850589433171125 \\
   0.194658379550856 \\
   0.049871016786613
    \end{pmatrix}, \\
    \ket{\psi_2}_{ABC} &=
    \begin{pmatrix}
    0.413378163566997 \\
  -0.732628740143910 \\
   0.202664162888498 \\
   0.302333182263915 \\
  -0.047229605421459 \\
  -0.088639894678373 \\
   0.345304731167627 \\
   0.174849887218713
    \end{pmatrix}, 
\end{align}
are written in the computational basis. From it, we consider the pair of two-qubit states from Eq.~\eqref{Eq:examples}, namely,
$\rho_{AB}\coloneqq{\rm tr}_C(\rho_{ABC})$ and $\rho_{AC}\coloneqq{\rm tr}_B(\rho_{ABC})$. As mentioned in the main text, we can verify that $\rho_{A} = \id/2$ and the minimum eigenvalues of the partial transposes of $\rho_{AB}$ and $\rho_{AC}$ are
$0.015170274157195$ and $0.000710511298615$, respectively.
Hence, both bipartite states are PPT~\cite{Peres1996PRL,Horodecki1996PLA} and therefore separable. Since $A$ is maximally mixed they can be viewed as Choi states of some entanglement-breaking channels.

Let $\rho_{BC}\coloneqq{\rm tr}_A(\rho_{ABC})$ be the two-qubit reduced state of $\rho_{ABC}$ [Eq.~\eqref{Eq:3-qubit example}] in $BC$.
Using the formula for the CHSH value of an arbitrary two-qubit state~\cite{horodecki2000PLA}, it can also be verified that 
\begin{align}
\mathrm{CHSH}(\rho_{BC}) = 2.048781741721893 >2,
\end{align}
meaning that $\rho_{BC}$ is nonlocal.
From the optimal CHSH observables for $\rho_{BC}$ we obtain the Bell operator:
\begin{equation}
\beta_{BC}^{\mathrm{CHSH}} \approx
\begin{pmatrix}
   -0.9860  & -1.5218  &  0.7221  & -0.4363 \\
   -1.5218  &  0.9860  & -0.4363  & -0.7221 \\
    0.7221  & -0.4363  &  0.9860  &  1.5218 \\
   -0.4363  & -0.7221  &  1.5218  & -0.9860
\end{pmatrix}. 
\end{equation}
We can use this as a linear witness for CHSH-nonlocality meta-transitivity for the $AB$ and $AC$ marginals of $\rho_{ABC}$. Using Eq.~(\ref{Eq: CHSHtransitivity}), we find that the compatible global state  $\eta_{ABC}$ with the smallest CHSH value in $BC$ gives
\begin{equation}
    \mathrm{CHSH}(\eta_{BC}) = 
    2.048514328822626 > 2.
\end{equation}
This demonstrates that nonlocality is unavoidable across all global extensions, and thus we conclude the proof of Result~\ref{Result 1}.
% \\

\subsection{Appendix B: Showing all joint broadcast extensions are non-entanglement-breaking}
% \noindent{\bf\em Showing all joint broadcast extensions are non-entanglement-breaking---}
To certify that all the broadcast channels realizing the entanglement-breaking marginal channels for our example in Result~\ref{Result 1} are not entanglement-breaking, it is sufficient to check if $\rho_{AB}$ and $\rho_{AC}$ have any global state $\eta_{ABC}$, which is possibly different from Eq.~\eqref{Eq:3-qubit example}, that has positive partial transpose in the bipartition $A|BC$. This is equivalent to checking if the following SDP has a non-negative optimal value:
\begin{align}
\label{Eq: PPTglobalSDP}
\nonumber \mu_\star :=
&\max_{\eta_{ABC}}\qquad\qquad \mu  \\
&\text{subj. to }  \quad
 \tr_{C}(\eta_{ABC}) = \rho_{AB},\,\,  
  % \nonumber\\  &\qquad \qquad \,\,\, 
 \tr_{B}(\eta_{ABC}) = \rho_{AC},\,\, \nonumber\\ 
 &\qquad\qquad\,\, \eta_{ABC} \succeq 0, \quad\eta_{ABC}^{\Gamma_A} \succeq  \mu \mathbb{I}, 
\end{align}
where $\Gamma_{A}$ denotes the partial transpose in $A$.
For the example in Result~\ref{Result 1}, we find
\begin{equation}
    \mu_\star \approx -0.068619998172138 < 0,
\end{equation}
which implies that there is no joint extension $\eta_{ABC}$ that has positive partial transpose with respect to $A|BC$.
Namely, all of them are entangled across the bipartition $A|BC$, meaning that they are all Choi states of some non-entanglement-breaking channels $A\to BC$.
% \\

\subsection{Appendix C: Output of the broadcast channel given Haar-random inputs}
% \noindent{\bf\em Output of the broadcast channel given Haar-random inputs---}
To study how well our compatible entanglement-breaking channels can produce entanglement and nonlocality with different inputs, we use the Choi state of their global extension given by Eq.~(\ref{Eq:3-qubit example}) to write down the action of the channel on an input state $\tau_A$ (below, the superscript $T$ denotes the transpose):
\begin{equation}
\label{Eq:broadcastChannelPureInput}
    \mathcal{G}_{A\to BC}(\tau_A) = 2 \tr_{A}\left( (\tau_{A}^T\otimes \id_{BC})\rho_{ABC}\right).
\end{equation}
Since we are interested in pure input states, it is convenient to write the input qubit in terms of its Bloch vector $\vec{r}$, which is then a unit vector that we can express in spherical coordinates. Let
$\tau_A = (\id + \vec{r}\cdot\vec{\sigma})/2.$

Then we have
$\tau_A^{T} = (\id + r_x\sigma_x - r_y \sigma_y + r_z\sigma_z)/2.
$
Substituting this back to Eq.~(\ref{Eq:broadcastChannelPureInput}), we get the following affine map for $\vec{r}$:
\begin{equation}
    \mathcal{G}_{A\to BC}(\vec{r}) = \rho_0 + \vec{r}\cdot\vec{\rho} =: \tau_{BC},
\end{equation}
where $\rho_0 = \tr_{A}(\rho_{ABC})$,
$\rho_x = \tr_{A}(\rho_{ABC}(\sigma_{x}\otimes\id_{BC} ))$,
$\rho_y = \tr_{A}(\rho_{ABC}(-\sigma_{y}\otimes\id_{BC} ))$, and
$\rho_z = \tr_{A}(\rho_{ABC}(\sigma_{z}\otimes\id_{BC} ))$.
Thus, we can check for the negativity~\cite{VidalWener2002_negativity} and CHSH value of the output state $\tau_{BC}$ for uniform random Bloch vectors $\vec{r}$.
% \\

% \noindent{\bf\em Acknowledgements---}
% \noindent{\bf ACKNOWLEDGEMENTS}
\section{ACKNOWLEDGEMENTS}
% \noindent 
The authors are grateful for helpful discussions with Antonio Ac\'in, Kai-Siang Chen, Yeong-Cherng Liang, Shiladitya Mal, and Paul Skrzypczyk.
C.-Y.~H. acknowledges support from ICFOstepstone (the Marie Sk\l odowska-Curie Co-fund GA665884), the Spanish MINECO (Severo Ochoa SEV-2015-0522), the Government of Spain (FIS2020-TRANQI and Severo Ochoa CEX2019-000910-S), Fundaci\'o Cellex, Fundaci\'o Mir-Puig, Generalitat de Catalunya (SGR1381 and CERCA Programme), the ERC Advanced Grant (CERQUTE and FLQuant), the AXA Chair in Quantum Information Science, the Royal Society through Enhanced Research Expenses (NFQI), and the Leverhulme Trust Early Career Fellowship (``Quantum complementarity: a novel resource for quantum science and technologies'' with Grant No.~ECF-2024-310).

\bibliography{Ref.bib}

\end{document}